\documentclass[
]{ceurart}

\usepackage[frozencache,cachedir=.]{minted}
\setminted{breaklines=true}

\begin{document}

\copyrightyear{2025}
\copyrightclause{Copyright for this paper by its authors. Use permitted under Creative
Commons License Attribution 4.0 International (CC BY 4.0).}
\conference{Developers Workshop, co-located with SEMANTiCS’25:
International Conference on Semantic Systems, September 3–5, 2025, Vienna, Austria}

\title{Jelly: a Fast and Convenient RDF Serialization Format}

\author[1,2]{Piotr Sowiński}[%
orcid=0000-0002-2543-9461,
email=piotr@neverblink.eu,
url=https://ostrzyciel.eu/,
]
\cormark[1]
\address[1]{NeverBlink, ul. Wspólna 56, 00-684 Warsaw, Poland}
\address[2]{Warsaw University of Technology, Pl. Politechniki 1, 00-661 Warsaw, Poland}

\author[1,2]{Karolina Bogacka}[%
orcid=0000-0002-7109-891X,
email=karolina@neverblink.eu,
]

\author[1,2]{Anastasiya Danilenka}[%
orcid=0000-0002-3080-0303,
email=anastasiya@neverblink.eu,
]

\author[1,2]{Nikita Kozlov}[%
orcid=0009-0003-1634-9194,
email=nikita@neverblink.eu,
]

\cortext[1]{Corresponding author.}

\begin{abstract}
    Existing RDF serialization formats such as Turtle, N-Quads, and JSON-LD are widely used for communication and storage in knowledge graph and Semantic Web applications. However, they suffer from limitations in performance, compression ratio, and lack of native support for RDF streams. To address these shortcomings, we introduce Jelly, a fast and convenient binary serialization format for RDF data that supports both batch and streaming use cases. Jelly is designed to maximize serialization throughput, reduce file size with lightweight streaming compression, and minimize compute resource usage. Built on Protocol Buffers, Jelly is easy to integrate with modern programming languages and RDF libraries. To maximize reusability, Jelly has an open protocol specification, open-source implementations in Java and Python integrated with popular RDF libraries, and a versatile command-line tool. To illustrate its usefulness, we outline concrete use cases where Jelly can provide tangible benefits. We consider that by combining practical usability with state-of-the-art efficiency, Jelly is an important contribution to the Semantic Web tool stack.
\end{abstract}

\begin{keywords}
  RDF \sep
  Serialization format \sep
  RDF stream processing \sep
  Knowledge graphs \sep
  Semantic Web data formats
\end{keywords}

\maketitle

\vspace{-0.2cm}
\section{Introduction}

Knowledge graph and Semantic Web systems use serialization formats such as Turtle, N-Quads, and JSON-LD to exchange and store RDF data in scenarios ranging from database dumps to client-server communication. However, common RDF formats are limited in terms of their serialization/deserialization speeds, compression ratios, and processing efficiency, which can become a performance bottleneck in real-life use cases (see Section~\ref{sec:usecases}). Additionally, no W3C format can natively represent streams of RDF data, which could be beneficial in applications such as industrial IoT, clickstreams, or real-time user interactions.
Although several new RDF formats were proposed in the past to solve some of these issues (see Section~\ref{sec:related-work}), none have reached wider adoption, either due to missing tooling, incomplete use case coverage, or other practical issues.

In this contribution we present \textbf{Jelly}, a high-performance binary RDF serialization format designed specifically to be easy to use, utilize as little compute resources as possible, and cover practical use case requirements. It natively supports RDF stream processing, but can also be applied in batch settings, in place of W3C-standard formats. In this work, we describe Jelly's contribution to the Semantic Web community tool stack that includes: (1) a robust and open Jelly protocol specification; (2) implementations for Java and Python compatible with popular RDF libraries; (3) an easy to use command-line tool; (4) several concrete use cases where Jelly can provide tangible benefits.

\vspace{-0.1cm}
\section{Jelly Protocol} \label{sec:protocol}

Jelly is a binary serialization format for streams of RDF triples, quads, graphs, or datasets~\cite{sowinski2022efficient}. It can be used in place of W3C-standard formats like N-Quads or Turtle, to simply represent a sequence of statements (a flat RDF stream in the RDF Stream Taxonomy~\cite{sowinski2024rdf}). It can also be used to represent a sequence of graphs or datasets (grouped RDF stream) -- this would be a series of files in W3C-standard formats. A Jelly file is split into \emph{frames}, where each frame corresponds to either a batch of RDF statements, or a complete RDF graph/dataset. In streaming settings (e.g., Kafka, gRPC), each frame corresponds to a separate message, allowing for efficient compression over the entire stream.

\paragraph{Serialization.} Jelly uses the industry-standard Protocol Buffers (Protobuf) serialization framework as its basis. The binary layout of Jelly is specified in a Protobuf interface definition file, which can then be used by compilers to generate serialization/deserialization code in one of the many supported programming languages (e.g., C++, Python, Java, Rust, C\#). This greatly simplifies implementing the format, as all binary-level manipulation code is generated automatically, and the developer only needs to implement a translation layer between Protobuf messages and RDF types of the given library. The mechanism for this translation is defined in the open Jelly specification\footnote{\url{https://w3id.org/jelly/dev/specification/serialization}}.

\paragraph{Compression.} The design goals of Jelly are to: (1) maximize serialization/deserialization throughput, (2) provide a relatively good compression ratio, and (3) operate in a fully streaming manner. We understand the third goal as meeting the common criteria for stream processing systems: process only one triple at a time, use a limited amount of time per triple, use a limited amount of memory (overall), output the results on demand, and adapt to temporal changes~\cite{bifet2023machine}. This essentially would allow Jelly to process arbitrarily large (potentially indefinite) RDF files in constant memory.

To meet these criteria, Jelly employs a streaming compression algorithm, defined in the aforementioned specification. The full description of the algorithm is beyond the scope of this contribution -- here we provide only a brief summary. Three fixed-sized string lookup tables indexed by integers are used, for IRI prefixes, suffixes, and datatypes. The serializer populates these tables until they are full, after which old lookup entries can be replaced by new ones using any policy (e.g., least recently used). This allows for processing datasets of indefinite size, as old lookup IDs can be reused. In triples and quads, IRIs are constructed as a pair of integer identifiers of the relevant prefix and suffix, referencing the lookup tables. These identifiers are represented as variable-length integers, to minimize their size. Additionally, simple delta compression is applied, with the value of zero having a special meaning (repeat last ID or increment it by one, depending on context). Because in Protobuf an integer with a value of zero is represented with zero bytes, this provides a very efficient compression mechanism. For consecutive RDF statements with repeating terms (e.g., same triple subject), the repeated term is not serialized, further reducing file size -- this is analogous to semicolons and colons in Turtle. These compression methods are relatively easy to implement and can work in a fully streaming manner, processing one triple at a time. They work to both reduce file size and to speed up processing, as less data written/read corresponds to less memory bandwidth, which is usually the constraining factor.

\paragraph{Performance.} We regularly publish performance benchmarks for the Jelly-JVM implementation on the Jelly website\footnote{\url{https://w3id.org/jelly/dev/performance}}, using a diverse mix of datasets from RiverBench~\cite{sowinski2024realizing}. In summary, Jelly-JVM 2.7.0 achieves an average compression ratio of 16.2\% (compared to a baseline of 100\% for N-Triples), serialization speed of 7.28 MT/s (millions of triples per second), and deserialization speed of 15.16 MT/s with Eclipse RDF4J. As far as we are aware, this currently makes Jelly the most compressed and the fastest RDF format implemented in either Apache Jena or RDF4J.

\section{Implementations and Tooling}

Jelly currently has implementations for Python and the Java Virtual Machine. Both implementations are designed around a generic core that can be used in conjunction with multiple different RDF libraries -- at the moment supported are: Apache Jena, Eclipse RDF4J, Titanium RDF API, and RDFLib.

\subsection{Java Virtual Machine: Jelly-JVM}

Jelly-JVM\footnote{\url{https://w3id.org/jelly/jelly-jvm}} is the more mature and heavily optimized implementation. It is organized into a set of modular components. The \texttt{jelly-core} module encapsulates the base functionality for encoding, decoding and transcoding Jelly data, independent of any particular RDF library, along with necessary utilities and abstractions to facilitate the development of integrations with various RDF libraries. The \texttt{jelly-jena} module provides full interoperability with Apache Jena by converting between Jelly Protobuf messages and Jena's data structures. It includes an integration with Jena's RIOT subsystem, making it possible to use Jelly in Jena just like any other RDF format (e.g., Turtle), including proper support for content negotiation. Module \texttt{jelly-rdf4j} offers analogous adapters for Eclipse RDF4J, in addition to an integration with the Rio serialization subsystem. Module \texttt{jelly-titanium-rdf-api} supplies an adapter layer for the minimalistic Titanium RDF API. Finally, the \texttt{jelly-pekko-stream} module written in Scala integrates with the powerful Apache Pekko Streams framework, allowing for efficient and scalable processing of RDF streams in more advanced use cases (e.g., IoT data aggregation).


Jelly-JVM can be used as a plugin with Jena or RDF4J\footnote{\url{https://w3id.org/jelly/jelly-jvm/dev/getting-started-plugins/}}. For example, for Apache Jena Fuseki, full Jelly support can be installed by placing the plugin JAR in the \texttt{extra/} directory of the Fuseki installation. This will enable full support in SPARQL UPDATE queries, the graph store protocol, and in the graphical user interface. Alternatively, for programmatic use, one can include either the \texttt{jelly-jena} (for Jena) or \texttt{jelly-rdf4j} (for RDF4J) Maven artifact by adding the following to pom.xml:

\begin{minted}{xml}
<dependency>
  <groupId>eu.neverblink.jelly</groupId>
  <artifactId>jelly-jena</artifactId>
  <version>3.4.0</version>
</dependency>
\end{minted}

This pulls in \texttt{jelly-core} and all necessary converters. With this dependency in place, Jelly support is registered with Jena or RDF4J automatically, enabling the use of Jelly as an RDF format in code. For example, one can load a Jelly‐serialized graph via Jena RIOT using:

\begin{minted}{java}
Model m = RDFDataMgr.loadModel("https://w3id.org/riverbench/v/2.1.0.jelly");
\end{minted}

To write this graph back to a Jelly file, one can call:

\begin{minted}{java}
RDFDataMgr.write(new FileOutputStream("m.jelly"), m, JellyLanguage.JELLY);
\end{minted}

Streaming serialization/deserialization with Jena, RDF4J, and Titanium is also possible, hooking into the iterator-like interfaces of these libraries, which allows for processing RDF data one statement at a time\footnote{\url{https://w3id.org/jelly/jelly-jvm/dev/user/rdf4j}}. Using the Pekko Streams integration, it is also possible to manipulate RDF data in more complex ways, e.g., batching a stream of quads into a stream of datasets\footnote{\url{https://w3id.org/jelly/jelly-jvm/dev/user/reactive}} and sending it to a Kafka topic.

\subsection{Python: pyjelly}

\texttt{pyjelly}\footnote{\url{https://github.com/Jelly-RDF/pyjelly}} is a Python implementation of Jelly, designed to make the benefits of the format accessible to the broad Python audience. Distributed through PyPI\footnote{\url{https://pypi.org/project/pyjelly/}}, it can be easily installed on all major operating systems (Linux, Windows, macOS) using standard Python package managers such as pip. To support users in getting started with pyjelly, a documentation suite is available\footnote{\url{https://w3id.org/jelly/pyjelly}} including usage examples, up-to-date functionality information, and an API reference.

Like Jelly-JVM, pyjelly also has a generic serialization core, on top of which integrations with other libraries are built. Currently, the popular RDFLib library is supported in this manner along with integrations for Neo4j and NetworkX, however, more integrations are planned. Serializing an RDFLib graph to the Jelly format is as simple as installing pyjelly with RDFLib support, e.g., \mintinline{shell}{pip install pyjelly[rdflib]}, and specifying the Jelly format during serialization. As shown below, this is as easy as using built-in RDFLib formats:

\begin{minted}{python}
from rdflib import Graph
g = Graph()
g.parse("https://www.w3.org/2013/N-TriplesTests/nt-syntax-subm-01.nt")
g.serialize(destination="triples.jelly", format="jelly")
\end{minted}

Similarly, a Jelly file can be parsed back to an RDFLib \mintinline{python}{Graph}:

\begin{minted}{python}
g = Graph()
g.parse("triples.jelly", format="jelly")
\end{minted}

\subsection{Command-Line Tool: jelly-cli}

To make it easier to use Jelly in production, development, and CI pipelines, we have developed a user-friendly command-line tool, \texttt{jelly-cli}\footnote{\url{https://github.com/Jelly-RDF/cli}}. This utility supports manipulating Jelly files without the need to write any code beyond a single terminal command. \texttt{jelly-cli} supports the adoption and growth of the Jelly ecosystem by lowering the barrier to entry and helps with common development tasks. 
Below we briefly showcase several commands supported by \texttt{jelly-cli} (version 0.5.1):

\begin{itemize}
  \item \textbf{Convert RDF to Jelly:}~~\mintinline{shell}{jelly-cli rdf to-jelly input.nq --to=out.jelly}\\
  This command converts RDF data written in any W3C-standard format into Jelly. Additional options allow for configuring the compression settings. 

  \item \textbf{Convert Jelly to RDF:}~~\mintinline{shell}{jelly-cli rdf from-jelly input.jelly --to=out.ttl}\\
  This command converts a Jelly file to a chosen RDF syntax. Supported formats include W3C-standard formats and a human-readable version of Jelly, useful for debugging (\texttt{jelly-text}).

  \item \textbf{Inspect a specific Jelly frame in human-readable binary form:}~~\mintinline{shell}{jelly-cli rdf from-jelly input.jelly --out-format=jelly-text --take-frames=3..5}\\
  Combined with \mintinline{shell}{jelly-text}, the \mintinline{shell}{--take-frames} flag allows extracting individual frames for direct inspection and debugging of parts of the Jelly file.

  \item \textbf{Merge and transcode Jelly files:}~~\mintinline{shell}{cat in1.jelly in2.jelly in3.jelly | jelly-cli rdf transcode --to=merged.jelly --opt.max-name-table-size=8192}\\
  The transcode command allows for merging multiple Jelly files into one, as well as recompressing the contents with new settings, using a very efficient transcoding algorithm.

  \item \textbf{Collect Jelly file statistics:}~~\mintinline{shell}{jelly-cli rdf inspect ./in.jelly --per-frame=true}\\
  Summarizes information about the Jelly file’s contents and used compression options. When run with \mintinline{shell}{--per-frame=true}, returns detailed statistics for each frame.

  \item \textbf{Validate a Jelly file:}~~\mintinline{shell}{jelly-cli rdf validate file.jelly}\\
  Validates the Jelly file and optionally compares its contents against a reference RDF file. This command can also verify whether specific compression options were used during serialization.
\end{itemize}

\section{Use Cases} \label{sec:usecases}

Jelly was designed to flexibly fit the requirements of many practical use cases, improving processing speed and reducing compute resource usage. Below we list the intended technical use cases, along with two examples of how Jelly is used in other projects.

\begin{itemize}
    \item \textbf{Client-server communication} -- for example, a frontend component can be efficiently linked to the backend. Jelly can reduce the latency between user input and the result, improving the user experience.
    \item \textbf{Inter-service communication} -- complex backend applications often consist of a network of microservices that must exchange RDF data (e.g., databases, cron job workers, analytics pipelines). Jelly can be integrated into existing APIs or with a complete gRPC stack, improving efficiency.
    \item \textbf{Database dumps and bulk loads} -- large RDF datasets can be quickly written and read with Jelly, while taking up less storage space. This can reduce the time needed for routine database maintenance tasks and help minimize infrastructure costs.
    \item \textbf{Streaming ingest} -- Jelly can eliminate ingestion bottlenecks in systems processing large amounts of streaming data, improving responsiveness, and allowing the system to scale to larger problems.
    \item \textbf{Database replication and change capture} -- changes to append-only databases can be recorded with the Jelly-RDF protocol described in Section~\ref{sec:protocol}. Additionally, add/delete operations with transaction support can be recorded using the Jelly-Patch format\footnote{\url{https://w3id.org/jelly/dev/specification/patch}}, based on RDF Patch.
\end{itemize}

\vspace{-0.5cm}
\paragraph{Example use case 1: Nanopublication Network.} The Nanopublication Network is a decentralized infrastructure for publishing and querying scientific knowledge using Semantic Web technologies~\cite{kuhn2021semantic}. It consists of various services (e.g., storage, querying) exchanging small RDF datasets -- nanopublications. This inter-service communication has proven to be a bottleneck, mainly due to the need for requesting each nanopublication individually which adds overhead. We switched over the communication to Jelly streams over HTTP, which reduced the time to retrieve 60,000 nanopubs from the server from over an hour to less than 4 seconds. This solution is now used in the live Nanopublication Network\footnote{\url{https://nanopub.net}}.

\vspace{-0.2cm}
\paragraph{Example use case 2: RiverBench.} RiverBench\footnote{\url{https://w3id.org/riverbench}} is a community-driven collaborative benchmark suite for RDF systems~\cite{sowinski2024realizing}. Each dataset in RiverBench is distributed as both a single N-Quads file (flat RDF stream) and an archive containing a sequence of TriG files (grouped RDF stream). We have added Jelly as a third option that can replace both of these distribution formats. Jelly natively supports streams of graphs/datasets, which simplifies parsing the file, removing the need to manipulate TAR archive entries. On top of that, Jelly is much faster to parse, reducing the time needed to set up a benchmark.

\section{Related Work} \label{sec:related-work}

Several efficient binary formats for streaming RDF data were proposed in the past, such as ERI~\cite{fernandez2014efficient} or S-HDT~\cite{hasemann2012shdt}. However, to the best of our knowledge, none have public implementations compatible with any modern RDF library, and therefore they are not usable in practice. The Jelly protocol uses the findings from these studies, but also focuses on broader tooling support and usability in general.

Both Apache Jena and Eclipse RDF4J have their own binary formats. In Jena, there are two nearly identical formats (one based on Apache Thrift, the other on Protobuf) that have no built-in compression and are largely analogous to N-Triples~\cite{web:jena-binary}. RDF4J has a compressed binary format that uses streamed dictionary compression (similar to Jelly) and a custom serialization framework~\cite{web:rdf4j-binary}. In our benchmarks, RDF4J Binary is by far the closest contender to Jelly, with similar compression ratios, 1.09--1.31x slower serialization, and 1.96--2.14x slower deserialization than Jelly-JVM 2.7.0\footnote{\url{https://w3id.org/jelly/dev/performance/rdf4j}}. However, none of these formats has a complete, up-to-date specification, and they only work within one ecosystem (either Jena or RDF4J), limiting their impact on the community. They also do not support streams of graphs/datasets (only streams of triples/quads).

There are also binary RDF formats with different design goals than Jelly, resulting in a very different set of trade-offs. CBOR-LD is a tightly compressed binary version of JSON-LD~\cite{web:cbor-ld}. It is primarily designed to be as compressed as possible (for use in, e.g., QR codes), while sacrificing ease of use (need to design compression contexts by hand) and serialization speed. CBOR-LD has an active community, with a tool stack similar to Jelly, e.g., the \texttt{ld-cli} tool. Finally, HDT~\cite{fernandez2013binary} is an indexed binary format that allows for direct querying of RDF files. It does not support streaming compression, as the entire contents of the graph must be known before writing the file. It is typically integrated with RDF libraries as a storage backend (not a serialization format), making it applicable for different use cases than Jelly.

\vspace{-0.25cm}
\section{Conclusion and Future Work}

In this contribution, we present Jelly, a high-performance RDF serialization format designed specifically to be fast, easy to use, and meet the practical requirements of many real-life use cases. It is currently integrated with several popular RDF libraries and has a CLI tool available to aid with production workloads, as well as testing and development.

We are constantly improving Jelly and its tooling. In the near future, we plan to: finalize protocol test cases for conformance testing of implementations, expand the feature set of the Python implementation, and integrate Jelly with commonly-used Semantic Web tools. Further plans include more implementations (e.g., JavaScript, Rust, Kotlin), new integrations (e.g., Pandas, KGTK), adding support for serializing SPARQL result sets, introducing new compression schemes, and temporal indexing.

\textbf{We invite the community to contribute to the Jelly protocol and its tooling}, with feature requests, bug reports, new integrations, documentation and more. More information on contributing can be found here: \textbf{\url{https://w3id.org/jelly/dev/contributing}}

This work was financially supported from the European Funds under the Sector Agnostic path of the Huge Thing Startup Booster program (project no. 0021/2025, program FENG.02.28-IP.02-0006/23).

\textbf{Declaration on Generative AI.} During the preparation of this work, the authors used ChatGPT in order to draft content. After using this tool, the authors reviewed and edited the content as needed and take full responsibility for the publication’s content.

\bibliography{bibliography}

\end{document}